\documentclass[journal]{IEEEtran}
\usepackage{amsmath,amsfonts}
\usepackage{algorithmic}
\usepackage{array}
\usepackage[caption=false,font=normalsize,labelfont=sf,textfont=sf]{subfig}
\usepackage{textcomp}
\usepackage{stfloats}
\usepackage{url}
\usepackage{verbatim}
\usepackage{graphicx}
\usepackage{algorithm}
\usepackage{cite}
\usepackage{acronym}
\usepackage{xcolor}
\usepackage{colortbl}
\usepackage[table]{xcolor}

\begin{document}
\bstctlcite{IEEEexample:BSTcontrol}
\title{Metasurfaces-Enabled Wave Computing for Future Wireless Systems: Opportunities and Challenges}

\author{Zahra Rahimian Omam,~\IEEEmembership{Member,~IEEE,} Hamidreza Taghvaee,~\IEEEmembership{Member,~IEEE,} Ali Araghi,~\IEEEmembership{Senior Member,~IEEE,}\\ María García-Fernandez,~\IEEEmembership{Member,~IEEE,} Guillermo Álvarez-Narciandi,~\IEEEmembership{Member,~IEEE,}\\ George C. Alexandropoulos,~\IEEEmembership{Senior Member,~IEEE,} Okan Yurduseven,~\IEEEmembership{Senior Member,~IEEE,}\\ and Mohsen Khalily,~\IEEEmembership{Senior Member,~IEEE}
\thanks{Z. R. Omam, H. Taghvaee, A. Araghi, and M. Khalily are with the Institute for Communication Systems, University of Surrey, GU2 7XH Guildford, UK (e-mails: \{z.rahimianomam, h.taghvaee, a.araghi, m.khalily\}@surrey.ac.uk).}
\thanks{M. García-Fernandez and G. Álvarez-Narciandi are with Universidad de Oviedo, 33203 Gijón, Spain (e-mails: \{garciafmaria, alvareznguillermo\}@uniovi.es).}
\thanks{G. C. Alexandropoulos is with the Department of Informatics and Telecommunications, National and Kapodistrian University of Athens, 16122 Athens, Greece and the Department of Electrical and Computer Engineering, University of Illinois Chicago, IL 60601, USA (e-mail: alexandg@di.uoa.gr).}
\thanks{O. Yurduseven is with the Centre for Wireless Innovation, Queen’s University Belfast, Belfast BT3 9DT, UK (e-mail: okan.yurduseven@qub.ac.uk).}
\vspace{-0.5cm}}
\maketitle

\begin{abstract}

The next generations of wireless networks are envisioned to integrate communications, sensing, and computing into a unified platform, demanding ultra-high data rates, submillisecond latency, and unprecedented energy efficiency. However, conventional digital processors face limitations in scalability, cost, and power consumption that hinder this vision. Wave computing, enabled by programmable metasurfaces, offers an alternative paradigm according to which signal processing operations are implemented in the domain of the propagation
of electromagnetic waves. This approach transforms metasurfaces from passive wavefront shapers into functional analog processors capable of executing tasks such as beamforming, sensing, imaging, and machine learning at the speed of light with minimal power consumption. This article provides an overview of metasurface-enabled wave computing, highlighting its fundamental principles and key application scenarios for future wireless systems, including integrated sensing and communications, artificial intelligence acceleration, over-the-air channel estimation, and computational electromagnetic imaging. Future research directions are outlined in response to the major open challenges of the technology, aiming to enable large scale deployment of wave computing in practical wireless networks. 
\end{abstract}

\section{Introduction} 

\IEEEPARstart{T}{he} rapidly evolving technology of programmable metasurfaces is expected to revolutionize the radio frequency fronts of next generations extremely large antenna-array-based transceivers as well as allow dynamic over-the-air manipulation of information-bearing signals \cite{GGJ2024_all}. These metasurfaces, typically composed of metallic patches or dielectric engravings, offer tunable responses to impinging electromagnetic (EM) waves, can be configured in planar or multilayered structures with sub-wavelength thickness, and can have inter-element spacings close to the sub-wavelength scale. In the wireless community, they are mainly known as reconfigurable intelligent surfaces (RISs), featuring diverse hardware architectures and functionalities (e.g., integrating sensing and communications (ISAC)) that can be programmed across time, frequency, and space \cite{RISsurvey2023_all}, and are thus recognized as a scalable, cost-effective, and energy-efficient physical-layer technology for next generations of wireless networks.

Future generations of wireless networks, starting with the upcoming sixth generation (6G), are expected to offer far more than substantially increased data rates and ultra-low latency, targeting the unification of communications, sensing, and computing into a single integrated platform  ~\cite{RISsurvey2023_all,GGJ2024_all}. 6G, in particular, is expected to support unprecedented requirements, including data rates approaching 1 terabit per second, latency on the order of a millisecond, and scalable energy-efficient operation across a diverse range of devices. Achieving this vision places, however, extraordinary demands on signal processing and computational resources. As of today, computing tasks are predominantly accomplished by digital processors, which in certain cases are specifically intended for computationally demanding applications (e.g., inference and super-resolution sensing). When it comes to lightweight devices, like those in the Internet of Things, such application-specific processors may constitute a prohibitive component
in terms of size, cost, and power consumption. In fact, the inherent limitations of general-purpose digital processors have, over the past decade, motivated the exploration of alternative computing paradigms, including graphics processing units (GPUs)and application-specific integrated circuits (ASICs), as well as approaches relying on quantum and wave-domain computing ~\cite{silva2014performing_all,yang2023reconfigurable_all,stylianopoulos2025over_all}. The latter computing paradigm, which is the topic of this article, capitalizes on the inherent properties of the over-the-air propagation of EM signals which can offer degrees of freedom for implementing diverse computing tasks (e.g., equation solving ~\cite{silva2014performing_all}, edge detection ~\cite{yang2023reconfigurable_all},  and edge inference ~\cite{stylianopoulos2025over_all}) at the speed of light with increased energy efficiency. Wave computing employs continuous analog signals, allowing for highly parallel processing capabilities and significant power consumption reduction. This feature of over-the-air computations contrasts sharply with traditional complex digital computing approaches that rely on discrete binary states.

Wave computing is mainly performed via metamaterials~\cite{silva2014performing_all,yang2023reconfigurable_all, stylianopoulos2025over_all}, whose responses can be dynamically engineered to enable precise manipulation of the properties of EM waves. Interestingly, the combination of the different states of a collection of metamaterials can enable wireless environments capable of directly performing various computing tasks. Among the first works dealing with metasurface-based wave computing, belongs~\cite{silva2014performing_all} that presented a metasurface combined with graded-index waveguides and multilayered slabs, which was designed to achieve a desired spatial Green’s function. In another direction, over-the-air wave domain computing eliminates the need for traditional signal processing components in wireless communication systems, such as analog beamforming/combining circuitry and respective digital processing units, as well as analog-to-digital converters, thus streamlining the computing process~\cite{yang2023reconfigurable_all}. In ~\cite{stylianopoulos2025over_all}, layers of programmable metasurfaces were optimized to realize neural network layers for edge inference purposes.

In this article, motivated by the potential of the multitude of recently available forms of programmable multifunctional metasurfaces~\cite{RISsurvey2023_all} (e.g., active/passive RISs, dynamic metasurface antennas (DMAs), simultaneous transmitting and reflecting RISs (STAR-RISs), and stacked intelligent metasurfaces (SIM)) for manipulating EM waves in wireless communication systems and the growing computational requirements for the 6G and beyond network components, we elaborate on the potential metasurface technologies for light-speed, highly parallelized, and hardware-compact wave-domain-based signal processing. The following Section~II introduces the core principles of wave computing with programmable metasurfaces and outlines two paradigms: interactive and non-interactive computing. It also presents two recent prototypes of metasurface-enabled wave computing. Section~III explores the key applications of smart wireless computing environments enabled by multifunctional computational metasurfaces. Open challenges and future research directions for this emerging paradigm of computing are discussed in Section~IV, while Section~V concludes the article with perspectives on the role of wave computing for future wireless networks.

\begin{figure*}[!t]
\centerline{\includegraphics[width=\textwidth]{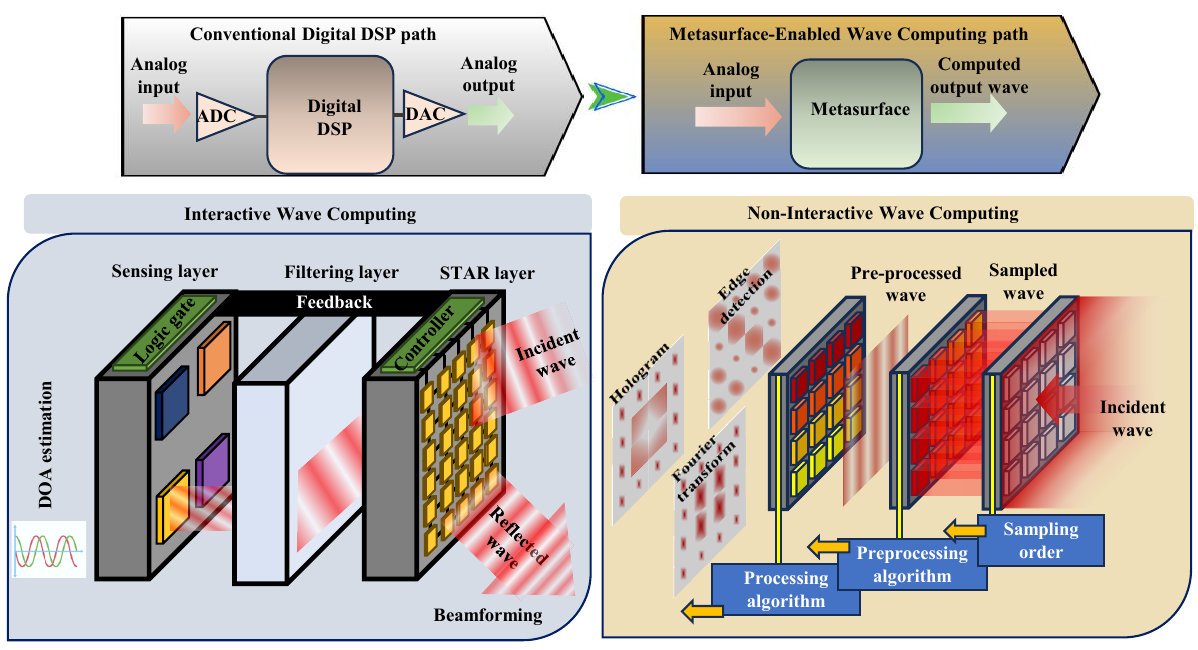}}
\caption{The main principle of wave computing with programmable metasurfaces: unlike digital signal processing (DSP), metasurfaces embed signal processing operations into their scattering profiles, transforming incident waves into processed outputs without need for digital conversion (top). Two paradigms of wave computing: interactive (bottom left), where sensing and feedback enable dynamic adaptation for tasks such as DOA estimation and beamforming; and non-interactive (bottom right), where predefined metasurface designs implement computations directly in the wave domain, such as edge detection or holography.}
\label{fig_1}
\end{figure*}

\begin{figure*}[!t]
\centerline{\includegraphics[width=\textwidth]{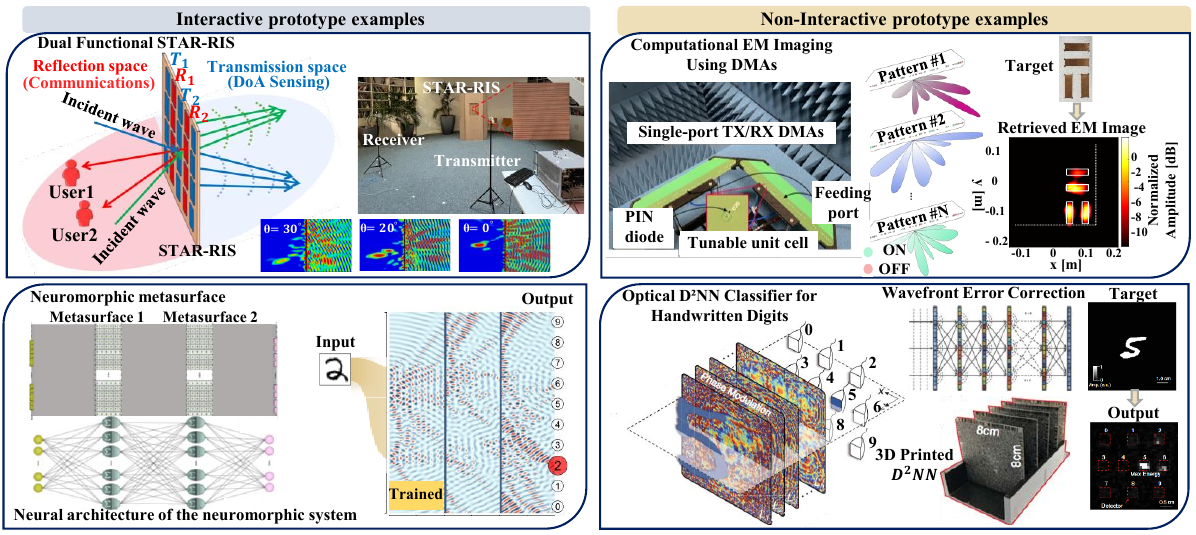}}
\caption{Representative Prototypes of Metasurface-Enabled Wave Computing. Interactive paradigms are demonstrated using (Top left) a dual-functional STAR-RIS that enables simultaneous wireless communication and real-time DoA estimation within a single aperture. (~\cite{omam2025star_all}), along with (Bottom left) a neuromorphic metasurface capable of task-driven analog inference (~\cite{mogh2024}). (Top right) Non-interactive paradigms include a DMA-based platform for computational EM imaging (~\cite{10564005_all}) and (Bottom right) a D²NN performing fixed-function optical classification and analog computation (~\cite{lin2018all_all}).\vspace{-0.2cm}}
\label{fig_2}
\end{figure*}




\section{Principles of Wave Computing}

Wave computing is a paradigm in which mathematical operations are carried out directly on physical waveforms as they propagate without requiring conversion into the digital domain. Unlike conventional digital processors that discretize signals into binary sequences, wave computing exploits the continuous properties of EM waves' amplitude, phase, frequency, and polarization to perform analog computations. Metasurfaces are central to this concept. By carefully engineering the geometry and material properties of their subwavelength elements, metasurfaces can embed mathematical operators into their scattering profiles. In doing so, they transform incident waves directly into computed outputs, effectively functioning as smart EM processors. This represents a shift from the traditional use of metasurfaces for wavefront shaping toward their role as multifunctional analog computing platforms. 



As illustrated in Fig.~\ref{fig_1}, two fundamental paradigms define metasurface-enabled wave computing: interactive and non-interactive. Interactive wave computing couples sensing, reconfigurable metasurfaces, and feedback control to achieve flexible and task-oriented processing. In this mode, incident signals are first captured and preprocessed by sensing and filtering layers, while a controller dynamically reprograms the metasurface response based on the acquired information. 
This closed-loop architecture allows real-time adaptation to environmental changes, supporting context-aware functionalities such as joint communication and direction-of-arrival (DoA) estimation using STAR-RISs, mobility-aware beamforming, and interference-driven spectrum shaping. Importantly the feedback loop operates on a separate and significantly slower control timescale and is not part of the computation itself. Instead, it updates the metasurface configuration in response to environmental changes, typically occurring on millisecond or longer timescales in practical 6G scenarios. This slower adaptation does not contradict the instantaneous nature of wave-domain computation. Once configured, wave computing proceeds purely through wave propagation, completing the computation within the subnanosecond aperture traversal time. It is worth emphasizing that, while both the proposed Interactive Wave Computing framework and the reconfigurable intelligent computational surfaces (RICS) \cite{yang2023reconfigurable_all} architecture aim to merge wave propagation and computation, the former emphasizes implementable, metasurface-based designs with greater hardware simplicity and integration feasibility, in contrast to RICS, which envisions multi-layer analog platforms incorporating neuromorphic and optical elements that remain largely conceptual. Non-interactive wave computing, in contrast, performs computations that are predefined and physically encoded in the metasurface design. Here, the scattering profile itself embeds mathematical operators such as Fourier transforms (FT), convolutions, differentiation, or edge detection. As a result of these operations, the sampled wave refers to the portion of the incident field selected or modulated by the metasurface for further computation, while the pre-processed wave denotes the transformed wavefront that has undergone analog operations (e.g., filtering or spatial coding) prior to detection. Once fabricated, the metasurface deterministically transforms incident waveforms into their computed outputs without requiring external control or adaptation. This paradigm is particularly attractive for applications where fixed high-speed analog preprocessing is sufficient, including real-time spectral analysis, analog feature extraction, and computational imaging. To further illustrate these paradigms, two prototype implementations from the previous work of the authors and two representative examples from the literature are presented.

\subsection{Interactive prototype examples}
A representative prototype of interactive wave computing is the design of a dual-functional STAR-RIS that integrates communication and computation within a single aperture, as shown in Fig.~\ref{fig_2} \cite{omam2025star_all}. At the core of its computational functionality lies the FT, a fundamental operator that bridges the spatial and frequency domains, with broad applications in signal processing, telecommunications, and imaging. In this architecture, an incoming wave is split into two paths: the reflected component is configured for non-specular beamforming to support communication, while the transmitted component performs near-field beam focusing, effectively implementing a discrete Fourier transform (DFT). This dual functionality enables simultaneous beamforming and DoA estimation. Unlike conventional computational RIS approaches that often require multilayered or metamaterial-based implementations, this STAR-RIS relies on a practical coding-based design that has been fabricated and experimentally validated. The result is a compact, efficient, and scalable platform that exemplifies how wave computing can directly augment 6G networks. Another interactive prototype is a neuromorphic metasurface system trained for task-specific analog inference, such as digit classification, using reconfigurable elastic layers \cite{mogh2024}. Once trained, the metasurface performs real-time wave-based inference without electronics, offering low-latency and energy-efficient edge intelligence.

\subsection{Non-Interactive prototype examples}
Another prototype involves computational EM imaging, where metasurfaces enable real-time, energy-efficient scene reconstruction. Such systems rely on raster scanning or large phased arrays, both of which are hardware- and power-intensive. In contrast, metasurface-based coded apertures generate spatiotemporally varying patterns that encode scene information into small numbers of measurements. In this context, the coded metasurface becomes part of the computational layer by synthesizing quasi-random fields, which serve as bases with minimized redundancy to probe the scene information and encode the backscattered data for computational imaging. A DMA provides a practical solution by using simple, low-power components (e.g., PIN diodes) to reconfigure the aperture. An example DMA-based single-pixel imaging system is shown in Fig.~\ref{fig_2}, where a bistatic aperture is dynamically reconfigured for data acquisition \cite{10564005_all}, providing high-resolution imaging with drastically fewer RF chains. \cite{lin2018all_all} demonstrates a complementary non-interactive paradigm at optical frequencies using diffractive deep neural networks (D²NNs). Here, a stack of trained diffractive layers performs fixed-function inference through passive wave propagation, enabling all-optical classification without electronics. This approach exemplifies metasurface-enabled computation in a non-reconfigurable setting, where the physical structure directly implements the neural-network function.

\begin{figure*}[!t]
\centerline{\includegraphics[width=\textwidth]{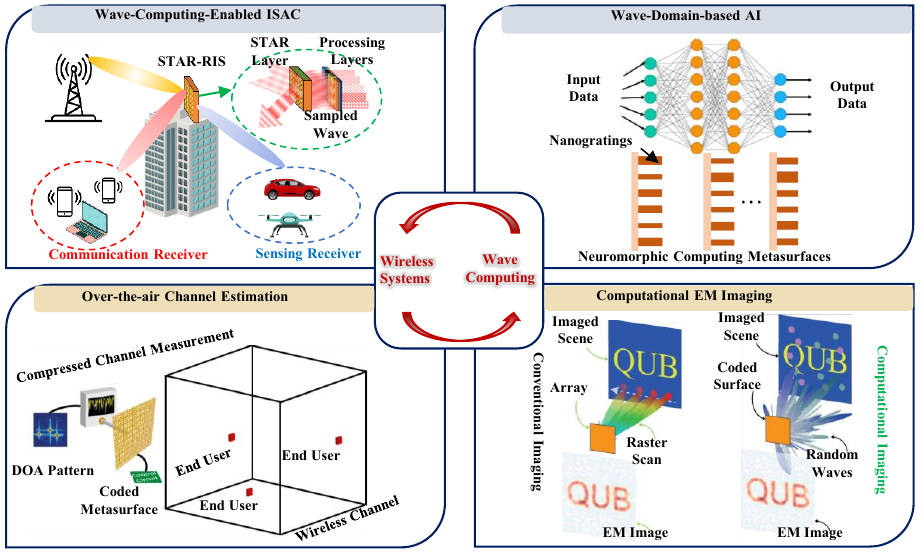}}
\caption{Wave computing in next-generation wireless systems, showcasing its application across diverse domains. (Top left) Wave-Computing-Enabled ISAC uses STAR-RIS to jointly serve communication and sensing by embedding processing into wave propagation. (Top right) Wave-Domain-Based AI employs neuromorphic computing metasurfaces, where nanostructured layers physically implement neural network computations for real-time inference. (Bottom left) Over-the-Air Channel Estimation leverages coded metasurfaces for compressed sensing and DOA estimation. (Bottom right) Computational EM Imaging replaces conventional raster-scanning with wave-based coding to reconstruct images using random scattering, enabling faster and more efficient imaging.
\vspace{-0.2cm}}
\label{fig_3}
\end{figure*}

\section{Wave Computing for Wireless Systems}
This section explores the transformative potential of wave computing in reshaping future communication systems for 6G and beyond, applications, as illustrated in Fig.~\ref{fig_3}. It introduces the essential components of a smart wireless computing environment, focusing on the pivotal role of computational multi-functional metasurfaces in enabling advanced communication and efficient processing capabilities.

\subsection{ISAC Enabled by Wave Computing}

Traditional signal processing methods for ISAC schemes and systems often involve digital techniques that, while precise, are computationally demanding and slow. These methods struggle to meet the ultra-low latency requirements of emerging 6G networks, where speed and efficiency are paramount. In contrast, wave-domain-based processing offers a compelling solution by performing complex mathematical transformations directly on the incoming wavefronts. This approach exploits the inherent parallelism of analog signals, allowing simultaneous processing across the entire wavefront. This capability promises a dramatic reduction in computational latency and a boost in processing speed, perfectly aligned with the latency demands of 6G applications \cite{yang2023reconfigurable_all}.

ISAC has been proposed to enhance wireless network performance by addressing the challenges associated with the unpredictable superposition of wireless signals. This technology aims to mitigate deleterious effects on RXs by integrating sensing capabilities, which allows for a dynamic understanding of the signal propagation environment. Wave computing can significantly enhance ISAC applications by enabling these systems to effectively decode superimposed signals from multiple sources or directions. By leveraging wave computing, ISAC systems can distinguish and process overlapping signals more efficiently, enhancing both the accuracy and reliability of combined sensing and communication operations. The ability to handle such superpositions in real-time further underscores the potential of wave-based processing to revolutionize ISAC systems, making it an indispensable component of future wireless networks. The analog nature of wave-based processing addresses a major challenge in ISAC functionalities by enabling precise manipulation of EM waves. As a practical example, by adjusting these properties directly at the RIS level~\cite{RISsurvey2023_all}, ISAC systems can achieve more refined control over the signal environment, enhancing both communication and sensing capabilities.

\subsection{Wave-Domain-based Artificial Intelligence}
As communication systems become increasingly complex, artificial intelligence (AI) appears as critical for managing dynamic networks, optimizing resources, and ensuring seamless performance. Neural networks, fundamental to AI, have advanced through progress in computer science and data availability, solving problems across communication networks, smart cities, and environmental monitoring. Modern AI models require massive computational resources, and running them at the network edge poses significant challenges. Wave computing offers a pathway to execute fundamental matrix operations, such as convolutions and Fourier transforms, directly in the electromagnetic domain. Such capabilities can accelerate inference for lightweight AI tasks in wireless nodes, reducing reliance on power-hungry GPUs and making real-time AI more practical in edge devices and base stations.

Wave computing will enable us to integrate physical systems into AI to perform computations in innovative ways, offering a potential solution to the aforementioned challenge. The integration of AI into metasurface-based wave computing, particularly through neuromorphic metasurfaces~\cite{mogh2024}, marks a transformative advancement in addressing the limitations of conventional digital computing in communication networks. Neuromorphic metasurfaces, which mimic the brain's neural structure and computation, enable efficient, real-time, and parallel data processing directly at the wave level. This ability to process information through wave interactions provides unprecedented speed, efficiency, and scalability, which are critical attributes for the demands of future communication systems. In this context, physical training will minimize the digital computation during training and indeed enhance the power efficiency. The physical local training algorithms, which are more compatible with broken-isomorphism physical neural networks, will support performing machine learning tasks with arbitrary analog devices.

\subsection{Wave-Domain-based Computational EM Imaging}
EM imaging is poised to become vital in future wireless networks by enabling real-time, energy-efficient sensing with simplified hardware. Metasurface-based wave-domain techniques, especially coded-aperture frameworks, are transforming traditional imaging by moving beyond conventional raster-scan methods~\cite{imani2020review_all}.
The coded metasurface becomes part of the computational layer, because the propagated waves essentially serve as random bases (with minimized redundancy) to probe the scene information and encode the backscattered data. Thus, the wave response of the coded metasurface is what enables the computational imaging. As a first step towards benefiting from wave-domain EM phenomena, a key enabler for the coded-aperture concept is the technology of metasurfaces. 
Leveraging metasurfaces for computational EM imaging can drastically reduce the number of data acquisition channels due to the single-pixel imaging approach offered by the coded-aperture concept in comparison to the conventional raster scan-based imaging modalities. An example DMA-based computational EM imaging system is shown in Fig. \ref{fig_2}, which forms a bistatic single-pixel aperture reconfigured using simple PIN diodes \cite{10564005_all}.



Despite the significant hardware-layer advantages that the coded-aperture-based computational EM imaging architectures offer, there are important challenges to be solved for these systems to serve as the technological backbone of 6G and beyond wireless networks. Computational imaging techniques relying on DMAs add an additional complexity to the signal processing layer due to the requirement to process large sensing matrices to solve the imaging problem \cite{7557020_all}. In this context, wave-domain-based computing to execute fundamental vector-matrix operations at the speed of light offers a significant potential to address the computational bottlenecks of conventional approaches and respective systems. In particular, metamaterial-based devices can be engineered to perform matrix inversions that can drastically reduce the processing latency to enable real-time imaging \cite{camacho2021single_all}. Analog wave computing can also play an important role in mapping wireless channels. Particularly, in the context of 6G, imaging tasks, such as identifying propagation paths with a minimal number of obstacles, will play a crucial role in offering reliable connectivity in cluttered environments. In this context, EM wave computing architectures can be used to enable post-processing on measured channel data and generate real-time EM maps. As an example, the first-order derivation of the EM channel maps to achieve edge-detection can offer a promising solution to identify and classify potential obstacles present in the wireless propagation environment \cite{wang2022single_all}.  
            
\subsection{Over-the-Air Channel Parameter Estimation}
Characterization of wireless channels is an important requirement for beamforming, which can be realized using metasurface-based apertures. With the ever-growing demands for next-generation wireless systems to offer reduced latencies and increased channel capacities, there is a significant need for technological breakthroughs, both in the hardware layer and the signal processing layer. In this context, from a channel perspective, increased carrier frequencies, such as the Frequency Range 2 (FR2) frequencies and the terahertz (THz) spectrum, can offer significant opportunities due to the availability of wider frequency bandwidths. Conventional channel characterization solutions rely on the use of array-based sounders to measure the channel response. A significant challenge with this approach is that such architectures can require an unpractically large number of elements, especially at high frequencies, due to the reduced wavelengths, as well as suffering from increased power consumption, circuit losses, and large costs of hardware components. RIS-based metasurfaces integrated with wave computing capabilities can offer advantages to tackle these challenges. 

Localization and DoA estimation play an important role in channel characterization tasks. Sensing the wireless channel and EM visualization (or mapping) of the channel will be one of the key ISAC enablers for future 6G networks~\cite{GGJ2024_all}. In this context, DoA estimation and channel sensing can be considered variants of an inverse problem, involving the retrieval of the location and DoA information of users. Solving these problems may require significant hardware resources due to the increased computational complexity of the processing algorithms, especially at high frequencies. Wave computing realized by metasurfaces can offer FT-based calculations executed at the speed of light to achieve spatial-frequency (or k-space) processing of the channel data \cite{cotrufoa2024metamaterials}. In addition, wave computing can also support efficient Channel Knowledge Map (CKM) construction through low-latency, analog spatial channel feature extraction. 

\section{Open Challenges and Future Trends}
Metasurface-enabled wave computing holds significant potential, yet several critical challenges remain unaddressed. 
Table~\ref{tab:combined_comparison} contrasts digital platforms with metasurface-based wave computing 
highlighting the latter's complementary strengths and the potential for hybrid architectures ~\cite{jouppi2017datacenter_all}.

\subsection{Open Challenges}
\subsubsection{Analog–Digital Integration} 

Wave computing offers ultra-low latency and high energy efficiency, but inherently limited precision and flexibility. Hybrid analog–digital computing architectures will be critical, where metasurfaces act as front-end preprocessors for high-throughput tasks, while digital processors handle precision-critical operations. Designing efficient interfaces for such integrations (ADC/DAC, control circuits, and feedback loops) is an open challenge. Data-efficient techniques like computational imaging, along with optimized analog–digital converters such as Successive Approximation Register (SAR) ADCs and time-interleaved SAR architectures, can help reduce power consumption by balancing resolution, speed, and energy use.

\begin{table*}[t]
\renewcommand{\arraystretch}{1.15}
\setlength{\tabcolsep}{4pt}
\caption{Comparison of Digital and Metasurface-Based Computing Approaches.}
\label{tab:combined_comparison}
\centering
\begin{tabular}{|>{\columncolor{gray!20}\bfseries}l|p{0.32\textwidth}|p{0.44\textwidth}|}
\hline
\rowcolor{gray!30}
\textbf{Aspect} & \textbf{Digital Computing (CPU/GPU/ASIC)} & \textbf{Metasurface-Based Wave Computing} \\
\hline
\textbf{Principle} & Digital processing after ADC/DAC; relies on discrete logic circuits & Direct analog operations (e.g., Fourier transform, convolution) embedded in wave propagation \\
\hline
\textbf{Latency} & Limited by conversion and electronic cycles (ns regime) & Near-instant, wave-speed processing (ps or below) \\
\hline
\textbf{Energy per MAC} & High to moderate (CPU: 10–100 nJ, GPU: 1–10 nJ, ASIC: 0.1–1 nJ) & Ultra-low (10–100 fJ) due to passive or weakly active operation \\
\hline
\textbf{Parallelism} & Requires multi-core or GPU-based architecture & Inherent via spatial wave interference \\
\hline
\textbf{Scalability} & Limited at mmWave/THz and by interconnect overhead & Naturally scalable, especially suited for high-frequency (THz) regimes \\
\hline
\textbf{Precision} & Arbitrarily high (e.g., 32/64-bit) & Limited by Signal-to-Noise Ratio (SNR) (typically $<$ 8-bit equivalent) \\
\hline
\textbf{Integration} & Mature CMOS-based ecosystems & Experimental; ongoing research in CMOS-compatible fabrication \\
\hline
\textbf{Hardware Architecture} & Centralized electronic hardware with digital logic & - Layered/cascaded architectures (e.g., RICS  ~\cite{yang2023reconfigurable_all}): Multi-layer, complex inter-layer alignment/coupling. 

- Single-aperture architectures ( authors’ prior work ~\cite{omam2025star_all}): 
Single-layer (STAR-RIS), integrated control, lower profile.\\
\hline
\textbf{Applications} & Channel estimation, baseband DSP, coding, general-purpose compute & Equalization, imaging, beamforming, smart wireless environments \\
\hline
\end{tabular}\vspace{-0.2cm}
\end{table*}


\subsubsection{Error Correction and Robustness}

Analog signals are inherently more vulnerable to noise and distortion due to their continuous nature. This susceptibility means that even minor variations induced by noise can significantly alter the information carried by signals. To ensure the integrity and reliability of data in analog form, error correction mechanisms must operate in real-time, adapting to continuous signal variations instantly. Implementing error correction at the hardware level in analog systems involves designing complex circuits capable of conducting necessary calculations and adjustments continuously. This not only helps in maintaining the fidelity of signal transmission but also enhances the overall system efficiency by reducing the need for frequent conversions between analog and digital formats. Optimized real-time error correction may be achieved through adaptive bitrate protocols, prioritized transmission of correction packets, and data caching to minimize redundant processing and reduce latency.

\subsubsection{Hardware Design}  
A key challenge is designing scalable and efficient hardware for metasurface-enabled wave computing. Even at low- to mid-mmWave frequencies, the large number of tunable elements (e.g., PIN diodes, varactors, MEMS) creates significant biasing and control overhead, which is exacerbated in stacked or multilayer architectures. Addressing this requires circuit- and system-level solutions such as hybrid architectures, simplified beamformers, optical feeding, or distributed control to enable real-time reconfiguration with low power consumption.


\subsubsection{Material Technology}  
Material innovation is essential for advancing metasurface-enabled wave computing, especially at mmWave and THz frequencies where conventional components face limits in speed, loss, and integration. Emerging materials like phase change media, graphene, and CMOS-compatible platforms offer pathways to overcome challenges. Achieving reliable, large-scale deployment will also require precision nanofabrication and advanced integration techniques.


\subsection{Future Directions}

\subsubsection{Goal-Oriented and Semantic Communications}
Future communication systems demand advancements in speed, bandwidth, and intelligent data processing to achieve efficiency. Goal-oriented communication ensures explicit results in applications like autonomous driving, where ultra-low latency and high reliability are essential. Similarly, semantic communication prioritizes the meaning and intent of messages, optimizing context-aware exchanges. Wave computing, powered by the continuous and simultaneous processing capabilities of analog systems enhanced by metamaterials, enables real-time data analysis and decision-making, adapting dynamically to high-efficiency demands. Wave-domain computing at the physical layer can have network-level benefits: it reduces medium access control (MAC)-layer latency through faster analog processing with minimal centralized coordination; it improves spectrum management via on-the-fly spectrum sensing and adaptive waveform transformations; and it enhances distributed network intelligence by allowing low-power, real-time analog preprocessing at edge nodes. Operating at the speed of light, it facilitates rapid semantic analysis with minimal latency, meeting complex data processing needs. By integrating wave computing into goal-oriented and semantic frameworks, networks can become more intelligent and adaptive, supporting dynamic environments and diverse requirements efficiently.

\subsubsection{Integration with Photonics and New Materials}
The fusion of metasurfaces with integrated photonic platforms promises ultrafast, broadband wave computing across optical and THz frequencies. Meanwhile, emerging materials such as graphene, phase-change media, and 2D semiconductors can offer tunability, nonlinearity, and memory effects, expanding the range of computable functions.

\subsubsection{Non-Local Metasurfaces and Beyond-Diagonal RIS Architectures}

Non-local metasurfaces and advanced RIS architectures offer enhanced computational capabilities by leveraging inter-element interactions rather than local control alone. This enables richer analog wave transformations, supporting complex real-time signal processing and AI-inspired functions. Emerging designs using multilayer structures and engineered coupling expand the degrees of freedom, paving the way for scalable, low-latency computing at wave speed.


\subsubsection{AI-Driven Programmability}
Future metasurfaces will integrate machine learning algorithms into their control loops, enabling real-time optimization of electromagnetic responses. AI-assisted configuration can allow metasurfaces to self-adapt to channel conditions, user mobility, and application demands, moving toward autonomous intelligent computing surfaces.

\section{Conclusion}
Wave computing, enabled by advancements in metamaterials and EM principles, offers a transformative approach for future wireless communication systems. By leveraging real-time, high-speed, and energy-efficient wave interactions, this paradigm promises to address critical wireless system demands, such as ultra-low latency, high data rates, and scalable transceiver hardware architectures. Key applications, including ISAC, EM imaging, and AI-driven orchestration, emphasize the potential of wave computing to redefine wireless networks. While challenges, such as precise fabrication of metamaterials and integration with digital systems remain, ongoing research on the field promises to overcome these barriers.



\end{document}